**Mantle Differentiation, Mixing and Interior-Exterior Exchange**

Takashi Nakagawa[1*], Charitra Jain[2], Diogo L. Lourenço[3], Maxim D. Ballmer[4], Paul J. Tackley[3]

[1]Emerging Media Initiative, Kanazawa University, Kanazawa 920-1192, Japan. Also at Department of Earth and Planetary System Science, Hiroshima University, Higashi-Hiroshima 739-8526, Japan

[2]GFZ German Research Center for Geosciences, Potsdam 14473, Germany

[3]Institute of Geophysics, Department of Earth Sciences, ETH Zurich, Zurich CH-8092, Switzerland

[4]Department of Earth Sciences, University College London, London, WC1E 6BT, United Kingdom

[*]Corresponding author (Email: takashi.geodynamics@gmail.com)

Accepted as Chapter 2.4 of "Archean Earth"

Word count and number of figures: 10491 words, 4 Figures

# 1. Introduction

## 1.1 Overview

We review the formation of large-scale heterogeneity in the solid Earth from the magma ocean phase to the present day, focusing on lower-mantle structure and evolution, as well as continental formation and its impact on interior-exterior volatile exchange. In terms of volatiles, we focus particularly on water, as constrained by the evolution of relative sea-level. The solidification of the magma ocean sets the "initial condition" for solid-state mantle structure and evolution, partitioning volatiles between the surface and interior, and differentiating major and trace elements between mantle reservoirs. A possibly long-lived basal magma ocean is a potential reservoir for iron and incompatible trace elements with important implications for the present-day structure of the lowermost mantle and the distribution of heat-producing elements. With emergent plate tectonics and mantle convection, the production of oceanic crust is the dominant differentiation mechanism, and may contribute to lower-mantle heterogeneity. In turn, convection acts to (partially) re-mix heterogeneity. Thus, the structure of the lower mantle records the early differentiation of our planet, as well as long-term (tectonic) evolution. The formation of continental crust and lithosphere has also an important influence on deep-mantle structure and composition, because the distribution of continents can control the scale of plate tectonics, and thus affect mantle convection patterns. While the formation of continental crust and lithosphere, as well as style of early-Earth tectonics remain controversial, scenarios can be tested by self-consistent long-term evolution models. The exchange of water between the surface and mantle also plays an important role in the dynamics of Earth's deep interior. By using relative sea-level change as a proxy for the interior-exterior exchange of water, it is possible to constrain the deep-mantle water cycle. Based on a comprehensive review of these topics, we conclude with several open questions that should be addressed by future studies.

## 1.2 Motivation: The big picture

Understanding the evolution of the deep mantle from the earliest times to the present-day is intriguing because it is connected to the various geological and paleontological events on the Earth, as has been pointed out by geological, geochemical and geodynamic approaches (e.g., Condie et al., 2016; Gerya, 2014; Korenaga et al., 2017; O'Neill et al., 2022). In particular, the large-scale heterogeneous structures in the present-day Earth are a result of long-term geodynamic evolution, including early differentiation associated with the magma ocean solidification (e.g., Abe, 1997; Solomatov, 2000; Elkins-Tanton, 2008; Ballmer et al., 2017a) and ongoing differentiation associated with oceanic and continental crust formation. The largest structures are the Large Low Velocity Provinces (LLVPs) found in global tomographic images (e.g., Garnero and McNamara, 2008; Koelemeijer et al., 2016; Richards et al., 2023) and the distribution of oceanic and continental crust caused by the plate tectonic-induced supercontinent cycles (e.g., Mitchell et al., 2021; Cao et al., 2021 and references therein).

The early differentiation of Earth's deep interior, which creates the initial large-scale heterogeneous structure, can also generate the atmosphere-ocean system on the Earth via outgassing associated with the solidification of the surface magma ocean, which releases water and other volatile components (carbon dioxide, etc.) from the deep interior to the surface (Abe and Matsui, 1988; Zahnle et al., 2007; Elkins-Tanton, 2008; Hamano et al., 2013; Bower et al., 2022). The ocean created by the early differentiation may be preserved until the present-day Earth due to the two-way interior-exterior exchange of water associated with plate tectonics in mantle convection, depending on the possible amount of water that can be transported in the early Earth (e.g., Nakagawa et al., 2018). These examples of big-picture questions are some first order issues regarding the differentiation, dynamics and evolution of Earth's deep interior from the Archean to the present-day, and its contributions to interior-exterior exchange.

## 1.2 Structure of this review

In this review, we address the formation of heterogeneous structure in the Earth from magma ocean solidification to the present-day, generating structures that are imaged using seismological methods and chemical heterogeneity that is inferred from geochemical analyses (Section 2). In Section 3, we discuss the importance of geodynamic processes on volatile exchange (particularly water) which can have an important influence on the surface environment (atmosphere, ocean, sea level). In Section 4, we discuss future directions and open questions related to differentiation, mixing and interior-exterior change in geodynamic modelling approaches, followed by a summary section (Section 5).

## 2.    Differentiation: Magma ocean, early mixing, and preservation of chemical heterogeneity in the mantle

 The solidification of the Earth's magma ocean had far-reaching consequences, shaping the initial conditions for mantle convection, thereby controlling the tectonic style in the Hadean/Archean. Indeed, geodynamics research has demonstrated that various surface tectonic styles can be stable for identical physical parameters, depending on the initial conditions (hysteresis) (e.g., Weller and Lenardic, 2012). Furthermore, outgassing of the magma ocean has led to the formation of the atmosphere-ocean system. This process was triggered by significant magmatic outgassing, releasing volatile components like water vapour and carbon dioxide (Abe and Matsui, 1988; Elkins-Tanton, 2008; Hamano et al., 2013; Bower et al., 2022). At depth, the crystallisation of a basal magma ocean (BMO) likely resulted in large-scale heterogeneous structure in the deep mantle (Labrosse et al., 2007; 2015; Maurice et al., 2017; Ballmer et al., 2017a; Laneuville et al., 2018; Boukaré et al., 2018). Indeed, two types of magma ocean need to be considered: a surface magma ocean and a basal magma ocean. Soon after the BMO is separated from the surface magma ocean, solid-state mantle convection initiates in the crystal (i.e., cumulate) pile, which grows to form the Earth's mantle (Solomatov and Stevenson; 1993a,b,c; Maurice et al., 2017; Ballmer et al., 2017a). While the surface magma ocean solidifies rather quickly (within ~1 Myr; LeBrun et al., 2013), the surface expression of interior solid-state dynamics remains unknown. Suggestions for Hadean (and Archean) tectonics range from a single-lid planet to the immediate initiation of plate tectonics (for reviews see e.g. Korenaga, 2013; Gerya, 2014; Lenardic, 2018). What is certain is that convective flow in the mantle and the tectonic regime of the early Earth played a crucial role in the creation and redistribution of heterogeneity within the mantle, and notably in the formation of continental crust. In the following sections, we review the solidification and outcome of Earth's surface and basal magma oceans and their roles on the redistribution of mantle heterogeneity and formation of continental crust.

## 2.1 Surface magma ocean

Terrestrial planets such as the Earth are expected to evolve through an episode (or multiple episodes) of large-scale melting very early in their history. These episodes are commonly known as magma oceans and can happen through a combination of different processes that can heat up a planet. The main ones are gravitational energy release during core formation (Flasar and Birch, 1973; Sasaki and Nakazawa, 1986), radioactive heating by short-lived isotopes (Mostefaoui et al., 2005; Urey, 1955), thermal blanketing due to the presence of an early atmosphere (Abe and Matsui, 1986; Genda and Abe, 2003; Hamano et al., 2013; Lebrun et al., 2013), and shock heating due to giant impacts (e.g., Nakajima and Stevenson, 2015; Nakajima et al., 2021). For the Earth, the proposed moon-forming giant impact and related core segregation likely provided sufficient energy for a deep and global magma ocean episode (Tonks and Melosh, 1993; Canup, 2012; Ćuk and Stewart, 2012; Nakajima and Stevenson, 2015; Lock et al., 2018). The solidification of this last major magma ocean on Earth played an essential role in creating the proto-atmosphere-ocean system and the heterogeneous mantle structure prior to the initiation of plate tectonics (e.g., Schaefer and Elkins-Tanton, 2018).

Evidence for magma oceans in the early evolution of terrestrial planets comes particularly from Mars and the Moon. The anorthositic lunar highlands have been interpreted as cumulate piles that are formed by the floatation of crystals in a deep magma ocean (e.g. Elkins-Tanton et al., 2011). Seismic evidence on Mars indicates the preservation of a deep molten silicate layer, or basal magma ocean (Samuel et al., 2021, 2023; Khan et al., 2023). All terrestrial planets and planetoids in the solar system provide evidence for efficient metal-silicate fractionation (core formation) that implies the presence of widespread silicate melting early in their histories.

The solidification of a magma ocean is controlled by the heat balance between convective heat transfer within the magma ocean and radiative heat removal at the surface (e.g., Abe, 1997). Numerous uncertainties remain regarding the rate of radiative heat transfer at the surface, including aspects such as the formation of a thin crust, and the existence and optical thickness of an early atmosphere (e.g.,

Lebrun et al., 2013; Hamano et al., 2013; Bower et al., 2018). Due to this, the solidification time-scale of the surface magma ocean is still quite uncertain, ranging from thousands (Solomatov, 2000; Miyazaki and Korenaga, 2019b) to millions of years (Abe, 1997; Lebrun et al., 2013; Marcq et al., 2017; Nikolaou et al., 2019). These different solidification time-scales stem from differing assumptions regarding the process of magma ocean solidification. Lower-bound timescales assume the absence of an atmosphere, with the magma at the surface radiating as a black-body (e.g. Solomatov, 2000), certainly a simplification at least in the late stages of magma-ocean evolution. Upper-bound estimates take into account the influence of a proto-atmosphere formed by the outgassing processes occurring during the solidification of a surface magma ocean (e.g., Abe and Matsui, 1988). The outgassing of the magma ocean during its solidification has been shown to have the potential to form the atmosphere-ocean system on Earth and other Earth-like planets (Abe and Matsui, 1988; Elkins-Tanton, 2008; Hamano et al., 2013; Bower et al., 2022).

Figure 1 shows an example of a modelling effort to simulate the solidification of a magma ocean using the program SPIDER (Simulating Planetary Interior Dynamics with Extreme Rheology, Bower et al., 2018; 2019), which indicates that the solidification timescale of a surface magma ocean would be around ~5 kyrs if no atmosphere is present and batch crystallisation happens (see below). This timescale is consistent with the predicted value associated with theoretical scaling relationships for the same conditions (Solomatov, 2000).

One of the most important and complex questions regarding the crystallisation of a magma ocean is whether the crystals continue to be suspended in the liquid, or whether they settle, being removed from the liquid magma ocean (e.g., Patocka et al., 2022). The former end member is called batch or equilibrium solidification, and the latter fractional solidification (e.g., Caracas et al., 2019). If no heat buffer from an atmosphere exists, solidification proceeds fast and batch crystallisation is expected (Solomatov and Stevenson, 1993b; Lebrun et al., 2013; Bower et al., 2018). If an insulating atmosphere, such as a steam atmosphere, is present and heat loss is less efficient, then fractional crystallisation may occur, particularly during the late stages of crystallisation (e.g., Solomatov and Stevenson, 1993c; Hamano et al., 2013; Xie et al., 2020), leading to the progressive enrichment of the magma ocean in incompatible elements, including Fe and heat-producing elements. Such a progressive enrichment of the magma ocean creates extreme heterogeneity across the cumulate pile, and hence across the initial solid mantle (e.g., Elkins-Tanton, 2008) with important implications for the planet's structure and geochemical evolution.

## 2.2 Basal magma ocean

The basal magma ocean (BMO) hypothesis was first put forward by Labrosse et al. (2007). The authors proposed that the magma ocean starts crystallising at mid-lower mantle depths due to the density crossover between solid and molten silicate in the deep mantle (Stixrude and Karki, 2005; Mosenfelder et al., 2007; Caracas et al., 2019). The molten layers above and below the already crystallised mantle would keep crystallising, contributing to the growth of this mid-mantle solid layer. However, the BMO cools much slower than the surface magma ocean, such that the cumulate layer grows much faster towards the surface than the core-mantle boundary. Recent studies on the density of silicate liquids and co-existing solids indicate that the initial thickness of the BMO should be 300~900 km in this now "classical" BMO scenario, depending on the mode of crystallisation (fractional vs. batch) of the global magma ocean (Caracas et al., 2019).

Once the surface magma ocean has fully crystallised, the isolated BMO continues to cool very slowly as heat is transported to the surface by solid-state mantle convection. Since slow cooling promotes fractional crystallisation (Solomatov and Stevenson, 1993b,c), the BMO becomes increasingly enriched, with iron, incompatible trace elements, noble gases, and heat-producing elements (HPEs) partitioning into the liquid. Therefore, the newly solidifying crystal/cumulate layers also become progressively iron-enriched, and after a substantial part of the BMO is solidified, cumulates become sufficiently dense to resist entrainment by the overlying convecting solid mantle. Labrosse et al. (2007) argue that dense cumulates therefore accumulate as piles that may correspond to seismically-detected Large-Low Velocity Provinces (LLVPs) today. Furthermore, it has been pointed out that some highly enriched patches of the BMO have remained (partially) molten until the present-day and are seismically detected as the Ultra-Low Velocity Zones (ULVZs) at the core-mantle boundary (Labrosse et al., 2007; Pachhai et al., 2022). In any case, alternative hypotheses for the nature, origin and composition of both ULVZs and LLVPs abound (see reviews in, e.g., Garnero and McNamara, 2008; Tackley, 2012; Yu and Garnero, 2018). For example, LLSVPs can be alternatively explained by accumulations of recycled oceanic crust, or may be even purely thermal in origin (e.g., Christensen and Hofmann, 1994; Brandenburg and van Keken, 2007; Davies et al., 2015; see below). ULVZ have been associated to a wide range of solid or partially liquid iron-rich materials at the core-mantle boundary, originating from poorly-constrained mantle heterogeneity and/or core-mantle interaction (Yu and Garnero, 2018; Jackson and Thomas, 2021).

### 2.2.1 Basal Magma Ocean formation scenarios

Since the seminal work by Labrosse et al. (2007), it has been argued that a BMO should be formed on Earth and most rocky planets, at least for Earth size and larger. For example, it has been also suggested for Venus (O'Rourke, 2020). Notably, seismic evidence indicates the presence of a ~150 km thick BMO in the lowermost mantle of present-day Mars (Samuel et al., 2023; Khan et al., 2023). Here, we review the formation of a BMO under various planetary accretion and differentiation scenarios.

It is still unclear if the crystallisation of the Earth's magma ocean started from the bottom or somewhere in the mid-mantle. For example, Thomas et al. (2012) demonstrated that bottom-up crystallisation should occur when assuming the chondritic liquidus of Andrault et al. (2011), while crystallisation from the mid-mantle is suggested when considering the peridotite liquidus by Fiquet et al. (2010). Caracas et al. (2019) predicted crystallisation from the mid-mantle by combining experimental and molecular dynamics calculations. Nevertheless, it has been argued that a dense, Fe-rich BMO should be formed independently of whether a melt of composition similar to that of the bulk mantle is more or less dense than crystals in equilibrium (Labrosse et al, 2015; Ballmer et al., 2017a, Laneuville et al., 2018).

In the event of a global magma ocean crystallising from the bottom up, fractional crystallisation causes the composition of both liquid and solid to evolve. First crystals formed in the lower mantle are predominantly bridgmanite (e.g., Ohtani, 1985; Caracas et al., 2019), leading to an initial solid lower mantle that is enriched in silicate (i.e., low Mg/Si~1.0). With progressive crystallisation, the liquid MO and co-existing crystal are progressively enriched in Fe. The progressive (upward) enrichment of solid cumulates in FeO promotes gravitational instability (Hess and Parmentier, 1995; Elkins-Tanton et al., 2005). During such a Rayleigh-Taylor overturn, potential energy is transformed into heat by viscous dissipation, and since all the solid is close to its melting point, a large part can re-melt partially. Due to partial melting, the resultant liquid should be Fe-rich, and usually denser than the solid at high pressures, hence accumulating at the base of the mantle, potentially forming a BMO (Ballmer et al., 2017a).

In other words, the overturn of surface-magma-ocean cumulates (Elkins-Tanton, 2008) can provide an alternative mechanism to form a "secondary" BMO. This is especially important in the case of fractional crystallisation of the global magma ocean *and* low pressures at its base (smaller than ~115 GPa; Caracas et al., 2019), where a primary BMO due to flotation of crystals at mid-mantle depth (as suggested by Labrosse et al., 2007) cannot form. Relatively low pressures at the base of the global magma ocean may be relevant for cases where the planet's silicate shell is not completely molten after the (final) giant impact, or if no notable giant impact occurs during the late stages of planetary accretion, as suggested for Venus (e.g., Salvador et al., 2023). In any case, for the Earth, most giant impact models predict the existence of a deep global magma ocean (Tonks and Melosh, 1993; Canup, 2012; Ćuk and Stewart, 2012; Nakajima and Stevenson, 2015; Lock et al., 2018). However, it is possible that pressures at the core-mantle boundary have not exceeded ~115 GPa due to ultra-fast spinning in the aftermath of the Moon-forming giant impact (e.g., Ćuk and Stewart, 2012; Canup, 2014; Lock et al., 2018). In this case, the Earth's BMO may have been mostly "secondary". A secondary BMO due to cumulate overturn has also been hypothesised for Mars (Samuel et al., 2021), for which strong iron enrichment (e.g. due to mostly fractional crystallisation of the surface magma ocean) is required to stabilise the observed present-day BMO gravitationally (Samuel et al., 2021). For Earth, hybrid primary-secondary scenarios for BMO formation during planetary growth are well conceivable, with intriguing implications for the storage of primordial (e.g., before the Moon-forming giant impact) geochemical signatures in the lowermost mantle.

Another mechanism for BMO formation is an initially superheated core. A very hot core may arise if, during core formation, most of the released gravitational potential energy partitions into the core (Solomon, 1979; Samuel et al., 2010; Rubie et al., 2015; Landeau et al., 2016) or possibly by high-energy late-accretion collisions (e.g., Marchi et al., 2023). A very high core temperature would make the bottom of the mantle melt. Fractional melting would generate a Fe-rich melt, which could be denser than the solid, thereby creating a BMO (Labrosse et al., 2015).

### 2.2.2    Some implications of a Basal Magma Ocean on Earth

The proposed BMO has significant implications for Earth's evolution from the Hadean to present-day. For instance, this layer likely exerted control and acted as a buffer for heat transport across the core-mantle boundary (CMB). Generating a geodynamo prior to inner core nucleation poses significant challenges (e.g., Nakagawa and Tackley, 2010; Olson, 2013), particularly if a global magma layer is present above the CMB (Monteux et al., 2016). If a BMO is present, the diminished heat flow across the CMB is unlikely to sustain the thermal convection in the core (prior to the inner core nucleation) necessary for maintaining a geodynamo. An additional power source for convection in the Earth's core becomes necessary, such as the proposed exsolution of light material across the CMB (O'Rourke and Stevenson, 2016; Hirose et al., 2017). Another proposed idea is that an early geodynamo could be generated within the BMO (Ziegler and Stegman, 2013; Blanc et al., 2020), as this molten region could be composed of iron-rich magma with high electrical conductivity (Stixrude et al., 2020).

As the BMO crystallises, first cumulates are expected to actively rise through the mantle, possibly contributing to the formation of the Archean continental crust (Wu et al.., 2023). During the later stages of progressive fractional crystallisation, the BMO is hypothesised to give rise to cumulate piles and small patches of residual highly-enriched melt, both of which would be seismically detectable as LLVPs and ULVZs, respectively (Labrosse et al., 2007; also see above). This intriguing hypothesis implies that remnants of the BMO have survived in the lowermost mantle and can be investigated geophysically and/or geochemically (as sampled by plumes; e.g., Mundl et al., 2017; Jackson et al., 2017). With improved constraints on the phase and melting relationships in the lowermost mantle (Boukaré et al., 2015; Miyazaki and Korenaga, 2019a; Nabiei et al., 2021), there is now an opportunity to evaluate this hypothesis.

Labrosse et al. (2007) suggested that cumulates from the late stages of progressive BMO fractional crystallisation may form piles with LLVP-like geometries. However, the density of the last 10% of the BMO cumulate sequence is predicted to be very high (~2,000 kg/m$^3$ denser than pyrolite) due to strongly iron-enriched compositions (Boukaré et al., 2015), inconsistent with LLVP geometry and density anomalies (e.g., Lau et al., 2016; Richards, 2023). For such strong intrinsic density anomalies, a global layer with little topography (instead of isolated piles) would persist in the lowermost mantle (e.g., Davaille et al., 1999; LeBars and Davaille, 2002; Nakagawa and Tackley, 2008). Future work will have to explore the conditions under which BMO cumulates (possibly mixed with recycled heterogeneity, see below) can form moderately iron-enriched piles consistent with seismic constraints (Vilella et al., 2021).

For end-member fractional crystallisation, the shrinking BMO would moreover have an extremely high concentration of heat-producing elements (HPEs), which would lead to very slow secular cooling, or even heating, of the BMO (such as suggested for Mars' BMO; Samuel et al., 2021;

2023). According to the phase diagram in Boukaré et al. (2015), a BMO of notable thickness should still be present for CMB temperatures above ~3,850 K (i.e., near the low end of estimates; Nomura et al., 2014). Thus, the implications of BMO fractional crystallisation are inconsistent with seismic observations for ULVZ total volume and geometry (e.g., Yu and Garnero, 2018). While it appears almost inevitable that a BMO existed in the early Earth (see section 2.1), the geophysically-constrained structure of the lowermost mantle points to the fact that it has *not* solidified due to end-member fractional crystallisation. Indeed, alternative scenarios need to be explored.

The formation, evolution, and solidification of the Earth's basal magma ocean raise numerous unresolved issues concerning the emergence of deep mantle heterogeneity (see more in section 2.3) and the generation of a magnetic field by an early geodynamo. While several insightful hypotheses have been proposed over the last years, their validation through computational geodynamics as well as geophysical and geochemical data remains essential for advancing our understanding of Earth formation and differentiation.

## 2.3 Heterogeneity in the deep mantle

As the magma ocean crystallises, the silicate mantle starts convecting in the solid-state, governing the thermochemical evolution of the planet by controlling global-scale rock and volatile cycling. There are two perspectives on material mixing during mantle convection. Looking backward from the present-day, the mantle has been efficiently stirred by sinking slabs for at least 2–3 billion years due to plate tectonics as the governing mode of surface deformation. As there is ample evidence for whole-mantle convection (e.g., Grand, 2002; van der Meer et al., 2010), the mantle is expected to be well-mixed on large scales (e.g., Ferrachat and Ricard, 2001). Intermediate-scale heterogeneity is consistently introduced at subduction zones, where a slab sandwich of mid-ocean ridge basalt and harzburgite sinks into the mantle. This view of the mantle can be referred to as the "recycled streaks model", i.e., a well-mixed mechanical mixture of mafic-to-ultramafic rock types on small-to-intermediate scales (e.g., Allegre and Turcotte, 1986; Phipps Morgan and Morgan, 1999; Xu et al., 2008, See Fig. 2a) - also known as a "marble cake" structure (Allegre and Turcotte, 1986).

On the other hand, looking at mantle evolution in chronological order leads to a different picture. From this perspective, mantle convection already initiates while the surficial magma ocean is still present. Freshly crystallised cumulates promote (small-scale) mantle overturns as they are typically unstably stratified (Ballmer et al., 2017a; Maurice et al., 2017; Boukaré et al., 2018). Fractional crystallisation of the magma ocean could lead to large-scale heterogeneity across the mantle (e.g., Elkins-Tanton, 2008) with density and/or viscosity contrasts that are potentially too strong to be removed by subsequent mantle convection (Garnero and McNamara, 2008; Ballmer et al., 2017b; Gülcher et al., 2020). Moreover, the solid mantle may co-evolve with a molten BMO for several billion years (see above), with additional opportunities for heterogeneity formation (Labrosse et al., 2007). This end-member of mantle mixing may be referred to as the "preserved blobs model".

Both the recycled-streaks and the preserved-blobs models make specific and testable predictions in terms of mantle evolution and present-day structure. These predictions are outlined below and can be tested by geophysical constraints as well as geochemical data.

In terms of geophysical observations, heterogeneity in the deep mantle is mainly evident as two large-scale slow seismic anomalies, referred to as the Large Low Velocity Provinces (LLVPs; e.g., Garnero and McNamara, 2008). These anomalies, located beneath Africa and the south-central Pacific, are several thousands of km wide and a few hundreds of km high, making up 3%~7% of the mantle's volume combined (Hernlund and Houser, 2008; Cottaar and Lekic, 2016). However, their composition and origin remain hotly debated.

Many studies suggest a thermochemical origin for LLVPs, interpreting the seismic anomalies as (meta-)stable piles of distinct material above the core-mantle boundary (e.g., Tackley, 1998; Davaille et al. 1999; McNamara and Zhong, 2005; Deschamps et al., 2012). The presence of such a major chemical anomaly in the lowermost mantle would have important implications for Earth's evolution and dynamics. A chemical anomaly is indeed supported by sharp seismic gradients at the edges of LLVPs (Ni & Helmberger, 2003; Sun et al., 2009; Frost and Rost, 2014), an anti-correlation of S-wave and bulk-sound anomalies as well as a high ratio between S-wave and P-wave anomalies in the lowermost mantle (Trampert et al., 2004; Tesoniero et al., 2016; Koelemeijer et al., 2016), and geodetic constraints for positive LLVP-density anomalies (Ishii and Tromp, 1999; Lau et al., 2017; Richards et al., 2023). The suggested compositions of these thermochemical piles range from recycled oceanic crust that segregates from the rest of the mantle to accumulate above the CMB (Christensen and Hofmann, 1994; Davies, 2002; Ogawa, 2003; Xie and Tackley, 2004ab; Brandenburg and van Keken, 2007; Jones et al., 2020) to various types of ancient/primordial heterogeneity (Labrosse et al., 2007; Deschamps et al., 2012; Zhang et al., 2016; Gu et al., 2016; Vilella et al., 2021; Yuan et al., 2023). Hybrid scenarios for pile composition with a primordial-recycled mixture (Tackley, 2012), or layering (Li et al., 2014; Nakagawa and Tackley, 2015; Ballmer et al., 2016), have also been suggested (referred to as "Basal Mélange", BAM; see Fig. 2b). All these potential LLVPs compositions are geodynamically feasible as long as the intrinsic density anomaly of LLVP materials is roughly 1.5–3% (e.g., Tackley, 1998; LeBars and Davaille, 2002; Tan and Gurnis, 2007; Tackley, 2015 and references therein), placing constraints on iron enrichment (and mineral composition). The BAM model is intriguing as purely basaltic/recycled LLVP compositions are difficult to reconcile with their geophysical anomalies (e.g.; Deschamps et al., 2012; Vilella et al., 2021; cf. Thomson et al., 2021), but subducted basalt is predicted by geodynamic models to readily segregate and accumulate as thermochemical piles (e.g., Brandenburg and van Keken, 2007; Nakagawa and Tackley, 2008; cf. Li et al., 2014).

Mantle upwellings may entrain LLVP materials and can be sampled at oceanic hotspots (Burke et al., 2008). Geochemical signatures of potentially LLVP-related ocean-island basalts indicate a major contribution from recycled oceanic crust and/or sediments (Jackson et al., 2018; Hofmann, 1997), but also point towards ancient or even primordial materials (Mundl et al., 2017; Jackson et al., 2017; Peters et al., 2018). These observations suggest that subducted oceanic crust can accumulate at the core-mantle boundary over billions of years, but also be recycled by upwelling mantle plumes. Modelling studies (e.g., Christensen and Hofmann, 1994; Davies, 2002; Xie and Tackley, 2004a,b; Jones et al., 2019) demonstrate that the residence times of trace elements in the mantle require recycled crust to be intermittently stored in a separate mantle reservoir, such as thermochemical piles. While some studies suggest that a purely thermal origin for LLVPs cannot be ruled out based on seismic constraints alone (e.g., Davies et al., 2015), geochemical and geodetic evidence indeed supports chemical heterogeneity in the lowermost mantle. In any case, constraints on the properties of LLVPs are insufficient to uniquely distinguish between a recycled and primordial origin, and thus between the recycled-streaks and preserved-blobs models.

Another prediction of the recycled-streaks model of mantle structure is that recycled basalt accumulates in the mantle transition zone (Christensen, 1997; Ogawa, 2003; Nakagawa and Buffett, 2005; Davies, 2008; Ballmer et al., 2015; Yan et al., 2020). Basalt is negatively buoyant throughout most of the mantle, except for the depth range of 660~720 km due to the effects of a series of phase transformations (Irifune and Ringwood, 1993). This density crossover sustains a "basalt filter": basalt delivered to the mantle transition zone (MTZ) tends to accumulate in the MTZ due to gravitational stability around 660 km depth. Conversely, harzburgitic rocks can accumulate just beneath the MTZ (Yan et al., 2020). Significant enhancement of the MTZ by basaltic, and of the uppermost lower mantle by harzburgitic rocks, is indeed indicated by seismic studies (Maguire et al., 2017; Yu et al., 2018; Wu et al., 2019; Bissig et al., 2022; Tauzin et al., 2022), supporting that recycled streaks make up much of the mantle.

As an argument in favour of the recycled-streaks model, it has been pointed out that the addition of oceanic lithosphere by subduction (and/or via dripping/delamination) would fully replenish the entire mantle's volume on a timescale that is significantly shorter than the age of the Earth (Phipps Morgan and Morgan, 1999). Using Phanerozoic subduction rates of 4 $km^2$/Myr (Matthews et al., 2016), this timescale can be estimated as ~2.4 billion years. Even though the onset age of plate tectonics as well as the efficiency of crustal recycling in the Archean (see below) remain debated, many studies agree that plate tectonics has been operating for at least 2–3 billion years (see Korenaga, 2013; Palin et al., 2020 for reviews). However, it remains unclear whether mantle mixing is pervasive and equally efficient across the lower vs. the upper mantle (e.g., Tackley, 2008). Moreover, if mantle flow remains mostly localised along narrow up-/downwelling conduits (e.g., Becker and Faccenna, 2011), either due to variable composition (see below) or rock fabric (Gülcher et al., 2022), mantle mixing would be critically delayed. Accordingly, pockets of primordial mantle may survive over the age of the Earth, within and/or beyond LLVPs, consistent with geochemical constraints (e.g., Mukhopadhyay, 2012; Caracausi et al., 2016; Mundl et al., 2017).

The preserved-blobs model posits that primordial materials can survive for the age of the Earth somewhere in the mantle (Becker et al., 1999), and the mid-lower mantle has been identified as a viable reservoir (Ballmer et al., 2017b; Gülcher et al., 2020; 2021). Delayed mixing of mantle heterogeneity is promoted by rheological heterogeneity, even at rather moderate intrinsic viscosity contrasts (Manga, 1996). In the lower mantle, the main minerals bridgmanite and ferropericlase display major viscosity contrasts, the former being likely much stronger than the latter (e.g, Girard et al., 2016; Tsujino et al., 2022; cf. Cordier et al., 2023). For any significant viscosity contrast between these minerals, possibly accentuated by faster grain growth in bridgmanite (Fei et al., 2023), even modest lateral variations in bridgmanite-to-ferropericlase ratio, and thus silica content, can lead to significant localization of deformation (Ballmer et al., 2017b; Gülcher et al., 2020; 2021). Such variations are expected to occur in the early mantle as long as any degree of liquid-solid fractionation occurs during fractional crystallisation of the deep magma ocean and/or BMO (see section 2.2.1). Consequently, bridgmanite-enriched regions in the early Earth tend to be inefficiently mixed such that blobs of variable size (or "Bridgmanite-Enriched Ancient Mantle Structures", BEAMS; See Fig. 2c) may survive. For example, large BEAMS can be preserved in the mid-mantle, as mantle convection and deformation and is localised around them, mostly along up-/downwelling conduits. Evidence for the preservation of intrinsically-strong domains in the mid-mantle includes plume deflection and slab stagnation at ~1000 km depth (Fukao and Obayashi, 2012; French and Romanowicz, 2015), broad horizontal seismic reflectors in a similar depth range away from major up-/down-wellings (Waszek et al., 2018), as well as a viscosity hill in the mid-mantle (Rudolph et al., 2015). Nevertheless, unequivocal evidence for BEAMS remains elusive, partly because of their neutral buoyancy and weak thermal anomaly, and thus seismic signature.

As an alternative physical mechanism, intrinsic density anomalies can lead to long-term storage of primordial materials in thermochemical piles or BAMs (see above). In any case, these two opportunities for primordial material preservation are not mutually exclusive (Gülcher et al., 2021; see Fig. 2d). During fractional crystallisation of the (basal) magma ocean, a series of cumulates is expected to be formed, ranging from intrinsically-strong bridgmanite to intrinsically-dense iron-enriched cumulates (e.g., Elkins-Tanton, 2008). While there are various viable physical mechanisms and ample geochemical evidence for the long-term preservation of primordial heterogeneity somewhere in the Earth's mantle, geophysical evidence remains inconclusive, particularly in terms of locating primordial/ancient reservoirs. On the other hand, crustal recycling has been going on for billions of years, very likely already before the onset of plate tectonics. Thus, the convecting part of the mantle, including most of the well-stirred upper mantle, is widely made up of recycled streaks, with accumulations of basalt in the MTZ (and lowermost mantle). A hybrid picture of the lower mantle emerges, possibly with primordial pockets within a recycled-streaks matrix, and with LLVPs that are made up of (ancient) recycled crust and/or a range of primordial lithologies.

## 2.4 Role of continents

Continents are large-scale heterogeneous structures floating atop the Earth's mantle, which have played an essential role in the evolution of the coupled core-mantle-ocean-atmosphere system during Earth's history (e.g., Cawood et al., 2022). Continental stability and longevity is attributed to their thick, buoyant and highly-viscous lithosphere, particularly in cratonic regions. Evidence from xenolith P-T data (Lee et al., 2011), observations of high seismic velocities (Jordan, 1979; Grand and Helmberger, 1984) and low surface heat flux (Pollack et al., 1993; Nyblade, 1999) suggest that cratonic roots are made of melt-depleted and dehydrated peridotites, which are highly viscous, and thus resistant to deformation in response to drag from the convecting mantle below (Hirth et al., 2000; Lee et al., 2011). Overlying this thick and cold lithosphere is a thin continental crust, which acts as a geochemical reservoir with its high concentration of incompatible heat-producing and trace elements (Hofmann, 1988). The following section gives a short overview on the proposed mechanisms and the efforts by the geodynamical community to model the formation and stabilisation of continents.

It is generally inferred that mantle differentiation from basaltic/mafic magma is required to generate felsic magma as the source of continental crust. For present-day conditions, the dominant role of convergent plate margins (subduction zones) over intra-plate tectonic settings (plume-associated magmatism) has been highlighted to source the basaltic protolith (Taylor and McLennan, 1985; Sun and McDonough, 1989; Rudnick, 1995; Arculus, 1999; Barth et al., 2000; Hawkesworth and Kemp, 2006). However, during the Archean Eon with its mantle potential temperatures suggested to be hotter than the present-day (Herzberg et al., 2010; Aulbach and Arndt, 2019), subduction processes might have been different (van Hunen and Moyen, 2012). The majority of the Archean continental crust that has survived until present-day is made of Tonalite-Trondhjemite-Granodiorite (TTG) rocks (Jahn et al., 1981; Drummond and Defant, 1990; Martin, 1994).

van Thienen et al. (2004) proposed a gravitational instability mechanism, in which the oceanic crust transitions into denser eclogite at a depth of 30 km, sinks into the mantle, and melts again to produce continental crust. Moore and Webb (2013) offered a heat-pipe mechanism for early Earth, in which volcanism dominates the surface heat transport resulting in the formation of a thick oceanic crust. In such a scenario, the bottom of this cold and thickened crust gets delaminated in the form of eclogitic drips, which have been demonstrated to generate felsic magmas (Johnson et al., 2013; Fischer and Gerya, 2016; Piccolo et al., 2020). Taking plutonic magmatism into account in their regional-scale models, Sizova et al. (2015) identified three tectonic settings, which could generate felsic magmas from hydrated oceanic crust: lower crustal delamination and dripping, small-scale overturns, and local thickening of basaltic crust. Rozel et al. (2017) argued for plutonism-dominated early Earth and showed the possibility of tracking TTG formation conditions (Moyen, 2011) in global-scale simulations. This led Jain et al. (2019) to model the self-consistent evolution of Archean continental crust (see Fig. 3) with results comparable to crustal growth curves proposed in the literature (see Dhuime et al., 2017; Korenaga 2018 and the references within).

Experimental studies have shown that water is a prerequisite for the formation of the Archean continental crust (Beard and Lofgren, 1991; Moyen and Stevens, 2006; Zhang et al., 2013). However, the tectonic settings that can bring both water and basaltic material to higher pressures (>1 GPa) where felsic magmas are produced remain debatable. In their numerical experiments, Roman and Arndt (2020) showed that magmatic intrusions make the lower oceanic crust dry and the authors advocated the role of subduction over delamination and dripping processes behind production of felsic magmas. Wu et al. (2023) offer an alternative mechanism to transport large amounts of water to the upper mantle, in which BMO gets water-enriched by progressive crystallisation, becomes gravitationally unstable and generates mantle overturns. Further geodynamical investigations are needed to test these scenarios coupled with the water cycle in a mantle-crust-ocean system.

Three separate hypotheses have been proposed as underlying mechanisms behind the formation and stabilisation of cratonic roots (Pearson and Wittig, 2008; Lee et al., 2011): melting above a large thermal plume (Herzberg, 1993; Griffin et al., 2003; Griffin and O'Reilly, 2007; Arndt et al., 2009), underthrusting of subducted oceanic lithosphere (Helmstaedt and Schulze, 1989; Canil, 2004, 2008; Simon et al., 2007; Pearson and Wittig, 2008), or accretion and thickening of already buoyant arc lithosphere (Şengör et al., 1993; Ducea and Saleeby, 1998; Kelemen et al., 1998; Parman, 2004). The reader is referred to Chapter 2.3 for further details.

Formation of cratonic roots has also been extensively modelled invoking a range of mechanisms: tectonic shortening followed by gravitational self-thickening (Wang et al., 2018), compression and thickening of cratonic lithosphere due to high stresses from plate tectonics (Beall et al., 2018), viscous underplating of low-density, melt-depleted sublithospheric oceanic mantle following subduction (Perchuk et al., 2020), melt-depletion induced dehydration stiffening (Capitanio et al., 2020), naturally occurring lithospheric compression and tectonic thickening (Jain et al., 2022).

Once formed, continents have a strong effect on mantle convection and thus deep mantle structure, as has been shown by a number of modelling studies. Because of their thickness and the enrichment of heat-producing elements in the continental crust, continents locally reduce heat loss from the mantle, causing the mantle below them to heat up. This may generate flow reversals under supercontinents, causing hot upwelling flow that breaks them up (Lowman and Jarvis, 1993), and inducing long-wavelength heterogeneity in the mantle (Gurnis and Zhong, 1991). It has been argued that the formation and disaggregation of supercontinents dominates Earth's tectonic evolution and produces an alternation between degree-1 and degree-2 mantle flow and structure (Li and Zhong, 2009), which strongly influences the evolution of dense provinces above the core-mantle boundary (Trim and Lowman, 2016). However, Rolf et al. (2014) found the cyclicity to be more statistical in nature. In any case, a concentration of stress tends to occur at continental edges, making them a preferred locale for subduction (Ulvrova et al., 2019) and even making plate tectonics more likely (Rolf and Tackley, 2011). The combination of continents and ocean plates leads to the triangular oceanic plate age-area distribution observed on Earth (Coltice et al., 2012). Continents reduce heat flow from the mantle below, increasing mantle temperature and influencing mantle thermal evolution (Grigné et al., 2007; Cooper et al., 2006, 2013), both because of their direct effect and because they contain a significant fraction of the global budget of heat-producing elements (Jaupart et al., 2015). The dynamics of continents and oceans has co-evolved with the state of the mantle over time, probably going through several distinct phases (Cawood et al., 2022).

## 3   Interior-exterior exchange: Impact of deep interior dynamics on ocean evolution

One of the most intriguing aspects of the evolution of the Earth since the Archean is why the Earth became a habitable planet, in contrast to other rocky planets in our solar system (Mercury, Venus and Mars). That is mainly because the Earth has a surface environment that allows for the existence of liquid water (ocean). Water vapour outgassing caused by magma ocean solidification permitted Earth to start with an ocean (Abe and Matsui, 1988; Zahnle et al., 2007; Elkins-Tanton, 2008; Hamano et al., 2013; Bower et al., 2022), and through material circulation caused by plate tectonics, the ocean has existed continuously from the Archean to the present day. In addition, preserving the warm climate so that the liquid water could remain as "ocean" over from Archean to the present is a significant factor (e.g., Kasting et al., 2006; Catling and Zahnle, 2020). Geological investigations have inferred the presence of oceans from at least 3 billion years ago to the present (Schubert and Reymer, 1985; Appel et al., 1998; Genda, 2016).

Sea-level change is a good proxy for constraining how the Earth's oceans have evolved over billions of years (e.g., Kasting et al., 2006; Karato et al., 2020) because it provides direct evidence of the time-variation of ocean water volume combined with secular variations in the ocean basin volume due to changing tectonics (e.g., Wright et al., 2020). In recent Earth history, for example, paleoenvironmental reconstructions indicate that sea level has decreased by nearly 100 metres during the last 65 million years (Miller et al., 2020).

The water cycle between the atmosphere and the ocean has a much shorter timescale (~100 years) than the geological timescale (~millions years) and is nearly balanced as a closed system with precipitation, evaporation and run-off (Trenberth et al., 2007; Bodnar et al., 2013). Therefore, long-term ocean volume variations are driven by water ingassing and outgassing. Outgassing of water is caused by volcanism. Ingassing occurs by seawater reacting with rocks in the oceanic crust and lithosphere and thus being integrated into the Earth's interior as water-bearing minerals, and much modelling has been performed to estimate how much water may be regassed into the interior this way (van Keken et al., 2011; Parai and Mukdopadhay, 2012; Cerpa et al. 2022). Water reduces the viscosity of mantle minerals as well as their solidus, thus increasing convective vigour, melting rate and outgassing rate. Thus, understanding the evolution of the coupled mantle-exterior system is of key importance for understanding Earth evolution.

Modelling of the coupled exterior-mantle long-term water cycle and the resulting evolution of ocean volume has typically been performed using parameterized ("box") models (McGovern and Schubert, 1989; Crowley et al., 2011; Sandu et al., 2011; Foley, 2015; Höning and Spohn, 2016; Seales and Lenardic, 2020), which provide a fast method of treating the various couplings and feedbacks and estimating the effect of key uncertain parameters and processes, including continental growth (Höning and Spohn, 2016) and exposed land area (Foley, 2015). However, the mantle cannot realistically be treated as a "box" with respect to water content because water solubility depends very strongly on mineral phase, temperature and pressure. This may result in a filtering mechanism that traps water in the mantle transition zone (Bercovici and Karato, 2003; Karato and Bercovici, 2006). Karato et al. (2020) examined the water filtering effect in more detail, in particular partial melting above the 410 km discontinuity, and argued that trapping of water in the mantle transition zone can help to explain the long-term trend of (little) sea-level change. In order to model such effects, global, multi-dimensional mantle convection models are required, but only a few such models have been published (Fujita and Ogawa, 2013; Nakagawa and Spiegelman, 2017; Nakagawa et al., 2018; Price et al., 2019; Nakagawa, 2023). Figure 4 shows the mantle water cycle and the relative sea-level change from the latest of these works (Nakagawa et al., 2023). The rate of relative sea-level change in the geodynamic model is consistent with paleontological observations (~1.0 m/Myr, Miller et al., 2020), but the time profile of sea-level change does not well match observations and geochemical modelling results (Parai and Mukdopadhay, 2012). The model did include water filtering above the mantle transition zone (Karato et al., 2020), which therefore seems insufficient for explaining the observed trend of long-term sea-level change. However, the implementation of the water filtering effect was simplified, so further investigations that include more detailed physics and petrology are required.

Thus, there is still a gap between observational data on relative sea-level change and estimates of the evolution of ocean water volume (which is used to evaluate sea-level change) based on the deep mantle water cycle. The reason is that the deep mantle water cycle involves complicated processes, including water-releasing metamorphic reactions that occur in and near subducting plates, and partial melting processes/regions in the Earth's deep interior. Additionally, the water storage capacity of the lower mantle is still quite uncertain because the behaviour of hydrogen in lower mantle minerals is controversial (e.g., Dong et al., 2021).

The combined evolution of ocean water volume and continental growth determines "continental freeboard", which is the relative elevation of the continental land masses with respect to sea level (Eriksson et al., 2006; Korenaga et al., 2017). The latter study employed principles of isostasy of oceanic and continental lithosphere (which have quite different density profiles) coupled with parameterizations of mantle evolution to estimate freeboard over the last 3.5 billion years. The evolution of relative sea-level is closely linked to the continental growth curve and does not vary greatly with time since, according to the models used, the amount of continental crust in the past was similar to the present-day value. However, there is currently little effort to perform computational geodynamical modelling of these aspects, and this is certainly needed.

Finally, the question arises of when and how much water existed on/in the Earth. This is still quite controversial because there are several hypotheses regarding how the water was transported into the early Earth (for a review see Genda, 2016). Using the D/H ratio to infer the amount of water on the early Earth (e.g., Kurokawa et al., 2018) leads to an estimate of 2 times the present-day ocean mass. Nakagawa et al. (2018) suggested that the total amount of water in the entire system of the early Earth was around ~5 to 10 ocean masses. During the solidification of the surface magma ocean, 1 to 5 ocean masses of water are needed in order to create the proto-ocean on the surface (e.g., Hamano et al., 2013). Therefore, up to 5 ocean masses of water should have been transported into the early Earth during the formation process.

## 4   Shortcomings and future directions

Mantle differentiation and mixing from magma-ocean crystallisation to the present day leads to the formation of large-scale heterogeneous structures in both surface and deep mantle regions, thereby affecting the style of mantle convection, and possibly even surface tectonics. In turn, the coupled Earth convective-tectonic system controls interior-exterior change of water as constrained by the evolution of relative sea-level. Below, we discuss the current state-of-the-art in terms of modelling this complex convective-tectonic-exterior system, and future directions for improving our current understanding.

## 4.1 Mantle differentiation and mixing

Mantle differentiation occurred both during the magma ocean phase, and in the subsequent long-term evolution to the present day. In section 2.3, two conceptual models are introduced: BEAMS, in which early-formed bridgmanite-enriched material persists to the present day due to high intrinsic viscosity, and BAM (basal mélange), which refers to long-term storage of dense material - a mixture of primordial material and recycled oceanic crust - at the base of the mantle. These concepts are not mutually exclusive, but can occur simultaneously, as shown by numerical models (Gülcher et al., 2021; Figure 2). However, this model must assume an initial condition that corresponds to the end of the magma ocean phase, rather than solving self-consistently for the entire Earth history. A future goal is to develop models that can model the transition between magma-ocean crystallisation and solid-state mantle convection.

The other main differentiation mechanism during Earth's history is the formation of continental crust, including the depletion of mafic components by the removal of melt, which may create TTG crust as the source rock of continental crust (see section 2.4). Recent years have seen tremendous progress in global mantle convection simulations (Rozel et al., 2017; Jain et al., 2019, 2022). However, the processes for (and timing of) continental crust production, the formation and stabilisation (over billions of years) of cratonic roots, and the transition from early Earth tectonic regime to subduction-driven plate tectonics remain poorly understood. A self-consistent model that can integrate all these aspects of Earth evolution, while at the same time satisfying geochemical constraints (isotopes and trace elements) and the geological record, is still lacking.

The large-scale heterogeneities discussed in this review, notably those near the thermal boundary layers (LLVPs, continents) influence the heat transport of mantle convection because they locally reduce heat transport across the CMB (e.g., Nakagawa, 2020; Nakagawa and Tackley, 2010) and surface (e.g., Grigné et al., 2007), respectively, particularly as being enriched in heat-producing elements. However, the initial state for the thermal evolution of the solid mantle, and thus the ideal initial condition for mantle-convection models, remains controversial. When parameterized convection models are run backwards in order to predict the thermal state of early Earth's mantle, they tend to lead to a "thermal catastrophe", in which the mantle temperature in the early Earth diverges to unrealistically high values (Davies, 1980). Proposed solutions to this include using an unrealistically high internal heating rate (e.g., Turcotte and Schubert, 2014), a gradual transition from layered to whole-mantle convection (Butler and Peltier, 2002), or an exponent in the Nusselt number-Rayleigh number scaling relationship that is much lower than conventionally assumed (e.g., Korenaga, 2006). However, Patočka et al. (2020) showed that allowing higher core heat flux than previously permits a reasonable thermal evolution using conventional scaling laws and whole-mantle convection.

In turn, 2D or 3D mantle-convection models with temperature-dependent viscosity and melting indicate that the present-day thermal state of the Earth's mantle does not depend significantly on the initial thermal state (e.g., Nakagawa and Tackley, 2010; 2012). In other words, the present-day mantle temperature can be naturally reconciled irrespective of the initial thermal state, and a "thermal catastrophe" can be readily avoided. In particular, the consideration of magmatic heat flux, which efficiently cools a planet that is "too hot" helps to avoid the thermal catastrophe but is usually neglected in parameterized convection models. In addition, the thermostat effect (i.e., negative feedback between mantle temperature and surface heat flux due to temperature-dependent rheology) regulates the long-term thermal evolution of rocky planets (e.g., Tozer, 1972).

In any case, it would be ideal to derive the initial state (both thermal and compositional) of the mantle from detailed forward models of magma ocean solidification, as discussed in sections 2.1-2.2. An important long-term goal is to develop a single simulation framework that can model self-consistently the solidification of the magma ocean, the transition to solid-state convection and long-term mantle evolution to the present-day. When such a model is successfully developed, it could address major questions in our field such as the style of Archean tectonics and mantle dynamics, as well as the onset of plate tectonics, the geodynamo, and inner core solidification. It could also address the heat flow across the core-mantle boundary over time, for which estimates for the present-day range from 8 to 20 TW (Lay et al., 2008; Frost et al., 2022). In order to resolve these issues, it would be required to develop more sophisticated modelling of the thermo-chemical evolution of Earth's mantle and core using numerical mantle convection simulations incorporating the formation mechanism of thermo-chemical structures caused by both the solidification of the magma ocean and subsequent differentiation.

## 4.2 Interior-exterior exchange

In section 3, we review the current state of knowledge and modelling of the Earth's deep long-term water cycle, which is caused by volcanic outgassing and ingassing due to subduction of hydrated rock, with relative sea-level change as the most important proxy (e.g., Karato et al., 2020). We conclude that while state-of-the-art 2D modelling (Nakagawa, 2023) (See Figure 4) can obtain a realistic amplitude of relative sea-level change and rate of sea-level change consistent with paleontological observations (~1.0 m/Myr), it cannot reproduce the long-term trend inferred from geochemical analyses (e.g., Parai and Mukhopadhyay, 2012). Reasons for this might be that (i) the geodynamical model does not include continental lithosphere, which has an important effect on sea-level and freeboard (e.g. Korenaga et al., 2017), (ii) water solubility in rock has a complex dependence on mineralogy, temperature and pressure and also influences melting behaviour, which may for example cause a transition zone water filter effect (Karato et al., 2020 and references therein), and this is currently simplified in geodynamical models, (iii) the initial/total amount of water in the Earth system is unknown.

So far, we have only addressed water cycling between the deep interior and surface. The long-term carbon cycle is perhaps equally important and intriguing because it strongly affects climate evolution from the Hadean to the present (e.g., McGovern and Schubert, 1989; Franck et al., 2002; Foley, 2015; Catling and Zahnle, 2020) which includes Snowball Earth events and the Great Oxidation Event, among others. There is currently much focus on the short-term global carbon cycle in the atmosphere-ocean system, due to global warming caused by human-related carbon emissions on a ~100-year time scale (e.g., Joos et al., 1999). This is different from the long-term carbon cycle that we discuss here, which involves volcanic outgassing of carbon dioxide (one of the main components of volcanic outgassing) (e.g., McKenzie et al., 2016) and carbon regassing into the interior by the subduction of carbonate rocks (e.g. Lee et al., 2019). The latter is thought to have caused a dramatic reduction in atmospheric carbon dioxide from Venus-like abundances in the Hadean to the small amount that exists today (Sleep and Zahnle, 2001; Franck et al., 2002; Zahnle et al., 2007; Catling and Zahnle, 2020). Understanding the deep carbon cycle is complicated because of the changing redox state in the Earth's deep interior, which controls the main component of volcanic outgassing (e.g., Bower et al., 2022). This is not yet fully understood from geochemical and mineral physics experiments (Frost and McCammon, 2008; Dasgupta and Hirschmann, 2010). In geodynamic modelling, investigations have been performed on the sensitivity of the redox state to outgassing (Guimond et al., 2021) and the carbon fluxes associated with recent plate tectonics (van der Meer et al., 2014; Wong et al., 2019; Müller et al., 2022). The mantle redox state regulates the amount of carbon dioxide outgassed (Guimond et al,, 2021) -- carbon dioxide outgassing is larger in an oxidised mantle - but more geodynamic modelling investigations are needed to study the deep carbon cycle on Earth because the behaviour of carbon in Earth's deep interior is complicated, as pointed out by geochemical and mineral physics experiments (e.g., Farsang et al., 2020).

## 4.3 Open questions

Below, we compile an (incomplete) list of key open questions that arise from the review above:

1. How does a BMO interact with mantle convection, from the Hadean to present? So far, BMO evolution models have been parameterized based on various assumptions (Labrosse et al., 2007, 2015; Laneuville et al., 2018; Blanc et al., 2020). It needs to be tested using detailed 2D or 3D long-term mantle convection models, how a BMO and ongoing convection and differentiation interact, and whether large-scale deep mantle heterogeneity is generated as a result. This has not yet been quantitatively addressed, because the timescales of convection in a magma-rich region vs. of the high-viscosity mantle are very different. Moreover, the style of BMO formation and subsequent crystallisation, which then may set up the extent of present-day deep-mantle heterogeneity (Labrosse et al., 2007), remains controversial (Ballmer et al., 2017a; Laneuville et al., 2018; Miyazaki and Korenaga, 2019b; Caracas et al., 2019; Wu et al., 2023).

2. What is the thermochemical structure of the lowermost mantle and how does it relate to (basal) magma-ocean crystallisation and/or the long-term tectonic evolution of our planet? In section 2.3, we discuss the nature and origin of LLVPs and ULVZs, which remains hotly debated. Both may be tightly related to early-Earth processes, core-mantle interaction, or the recycling of oceanic crust (and thereby the evolving tectonic style at the Earth's surface). Their (long-term) storage implies the intriguing possibility to test scenarios of (early) Earth evolution, even though the geologic record on our planet has been reworked by tectonic activity. Both these geophysical anomalies are also intriguing geochemical reservoirs that are possibly sampled by plumes. Another such proposed lower-mantle structure are BEAMS, the formation and preservation of which has been shown to be plausible, with some evidence suggesting their presence, but with unequivocal evidence still lacking. Better constraints on LLSVP/ULVZ composition, e.g. by ruling out some of the many proposed scenarios, as well as quantitative geophysical tests of the BEAMS hypothesis, are needed to improve our understanding of Earth evolution.

3. What is the heat flow across the core-mantle boundary? This is one of the most important quantities for long-term evolution of the deep mantle, and the mantle in general (e.g., Patocka et al., 2020). This quantity is still under debate, with current estimates ranging from 8 to 20 TW at the present day (Lay et al., 2008; Patocka et al., 2020; Frost et al., 2022). The heat flow across the core-mantle boundary is also reduced by the presence of dense material above the CMB (e.g., Nakagawa and Tackley, 2005, 2010, 2012, Nakagawa, 2020). The relevant constraints on the long-term evolution in the deep mantle are the continuous generation of the geomagnetic field observed over 4 billion years (Tarduno et al., 2015) and the age of the inner core, with estimates ranging from 0.5 to 2.5 billion years old (Biggin et al., 2015; Li et al., 2023). A complete computer simulation that satisfies both constraints has not yet been performed because (i) the thermal and compositional initial conditions for long-term mantle convection simulations are not yet well established (see Section 2), (ii) the effect of continental lithosphere has not been included - this is crucial in long-term mantle evolution due to affecting the efficiency of heat transfer (e.g., Cooper et al., 2013) and mantle convection pattern (e.g., Rolf et al., 2014).

4. What was the initial amount of water on/in Earth, and how was it distributed between surface, mantle and core? In an earlier study (Nakagawa and Spiegelman, 2017), this was not found to strongly affect the amount of water transported from the ocean to Earth's deep interior. However, for accurate modelling of the deep mantle water cycle that is consistent with the present-day ocean mass, the initial amount of water estimated from planetary formation studies (observational, theoretical, numerical and experimental) is an essential constraint.

5. How did the mantle redox state evolve with time, and how did this affect (and how was it influenced by) the long-term carbon and water cycles? The redox state of the deep mantle is typically quantified by the oxygen fugacity of the deep mantle minerals (e.g., Frost and McCannon, 2008). This is thought to have started relatively reduced and become more oxidised with time (Lecuyer and Ricard, 1999). Kuwahara and Nakada (2023) recently pointed out that the deep mantle may have reached an oxidised state in about 3 billion years. In computational geodynamics modelling it is presently difficult to include the oxygen fugacity, making it difficult to investigate such changes, something that must be improved in the future. Moreover, the carbon cycle could also be controlled by the carbonate solubility in the mantle rocks (Farsang et al., 2020). More work on the carbon cycle, along the lines presently done for water cycle modelling (e.g., Nakagawa, 2023) should be performed in the future.

6. How did the long-term deep interior-atmosphere carbon cycle influence Earth's climate over the whole Earth history? As discussed in Section 4.2, the carbon content of the atmosphere is thought to have changed enormously since the Hadean, and since carbon dioxide is a strong greenhouse gas, this will have affected the climate. Linked to this are the Great Oxidation Event, which may have been caused by a change in outgassing due to the redox evolution of Earth's mantle (e.g., Lee et al., 2019), and Snowball Earth events (e.g., Hoffman et al., 1998).

Elucidating the potential role of mantle geodynamics in these events is an important issue because there is currently a debate as to whether deep interior or biological mechanisms are more important.

7. How can we find an explanation for the isotopic geochemical observations in terms of mantle differentiation (e.g., melting) and mixing mechanisms? Differentiation has first-order effects on trace elements and thus radiogenic isotopic systems, because it fractionates parent and daughter isotopes, leading to isotope ratios that then evolve differently with time in different mantle materials. Mantle mixing competes with differentiation in terms of geochemical anomalies, but the timing of both processes is relevant for isotopic systems. The efficiency of mantle mixing is controlled by the physical properties (density, viscosity) of mantle rock types (section 2.3), which in turn are controlled by their major-element geochemistry (i.e., also affected by differentiation). Various geochemical observations are interpreted to provide evidence for primordial differentiation, ongoing differentiation and mixing, the addition of material after Earth formation ("late veneer"), for core-mantle reactions, and more. Progress has been made in modelling the evolution of isotopic systems in fully-coupled geodynamic models, but open questions remain. For a discussion of these issues, see e.g. recent reviews (Stracke, 2012; Tackley, 2015; Weis et al., 2023).

Addressing these issues will help to further our understanding of heterogeneity generation, mantle mixing, and interior-exterior exchange.

## 5   Summary

Mantle differentiation starts with magma ocean crystallisation, which includes a relatively short-lived surface magma ocean and quite likely a basal magma ocean that survives through much of the subsequent long-term evolution. During long-term evolution, the dominant differentiation mechanism is basaltic crustal production and subduction into the deep mantle (or before plate tectonics started, dripping and delamination), with the production of continental crust and lithosphere also being a key component. At the same time, mantle convection acts to mix the differentiated materials. Two conceptual models for the long-term preservation of heterogeneity are reviewed: the Basal Mélange (BAM), which is a mixture of primordial and/or recycled products accumulating above the core-mantle boundary due to their intrinsic density and seismically imaged as LLVPs, and Bridgmanite Enriched Ancient Structures (BEAMS), which originate in the lower mantle during magma ocean crystallisation and might survive over geological time due to high intrinsic viscosity of bridgmanite. These structures can occur simultaneously, as indicated by recent numerical modelling (Gülcher et al., 2021). The formation of early continental TTG crust by hydrous partial melting of basalt in non-subduction settings has also successfully been modelled in global geodynamical models (e.g., Jain et al., 2019). Geodynamic numerical modelling is thus a powerful approach for testing various hypotheses and combining mineral physics, geochemical and seismological constraints. However, a limitation of such modelling studies is that they start after magma ocean crystallisation and therefore the initial condition is not fully realistic or self-consistent. Developing models that can model from magma ocean to the present day is a goal for future investigations and would allow fully self-consistent thermo-chemical structures to be developed.

Interior-exterior exchange of water by volcanic outgassing and subduction of hydrated rock drives the long-term water cycle, which has important influences on both interior dynamics (water reduces viscosity and influences melting behaviour) and the surface environment (ocean volume, continental freeboard, life). While most modelling to date has used parameterized ("box") models, these may be inadequate because water solubility in the mantle is highly heterogeneous. It is now possible to perform multidimensional geodynamic modelling of water circulation in the mantle that incorporates detailed solubility maps as a function of temperature, pressure and composition and is thus able to include dehydration reactions along subducting plates, melting above the 410 km discontinuity, etc. Relative sea-level change is the most useful proxy of interior-exterior water exchange, but converting ocean volume to sea level requires including continents, and this has not yet been done in geodynamic models with water cycling, thus indicating an important direction for future modelling.

The long-term carbon cycle between Earth's deep interior and atmosphere is also important and intriguing because it strongly affects the climate evolution over billions of years. This is so far only been treated using parameterised models (e.g, Krissasen-Totton et al., 2018). Future investigations using multi-dimensional geodynamical models are needed to reveal how the dynamics of Earth's deep interior contribute to long-term climate evolution, including its role in the Great Oxidation Event, snowball Earth events, etc.

**Acknowledgements.**

T.N. thanks Shun-ichiro Karato and Masanori Kameyama for fruitful discussions on the interior-exterior exchange caused by mantle dynamics. C.J. is receiving funding from the ERC Synergy Grant 856555 for the project: Monitoring Earth Evolution through Time (MEET). We thank Anna Gülcher for providing the original version of figures that are part of Figure 2.

**Author's contribution.**

Takashi Nakagawa - Lead the entire plan, writing the original manuscript (section 1, 3, 4 and 5) and reviewing and editing of the entire document. Charitra Jain - Writing the original manuscript (section 2.4 and section 4.1). Diogo L. Lourenço - Writing the original manuscript (section 2.1 and 2.2). Maxim D. Ballmer - Writing the original manuscript (section 2.3 and section 4.1) and reviewing and editing of the entire document. Paul J. Tackley - Reviewing and editing of the entire document; integration, filling in gaps.

**Figures and captions**

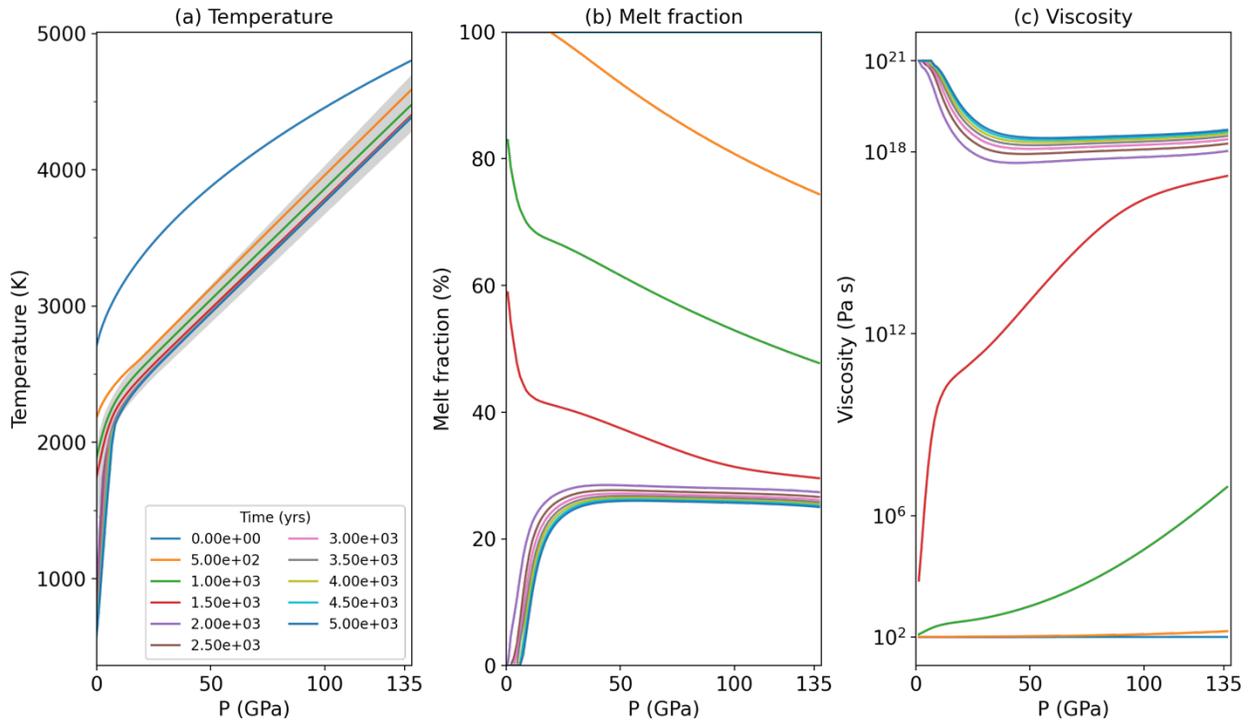

Figure 1. Time-evolution of (a) Temperature, (b) melt fraction and (c) viscosity of a magma ocean as predicted by 1D convection models based on mixing-length theory (Bower et al., 2018; 2019). The temperature range between solidus and liquidus in (a) is shaded.

**(a) Recycled streak model (Xu et al., 2008)**

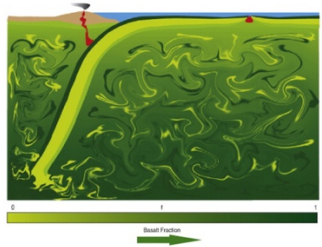

**(d) Reproduction of (a) to (c) in one numerical modeling (Gülcher et al., 2021)**

**(b) Basal Mélange (Tackley, 2012)**

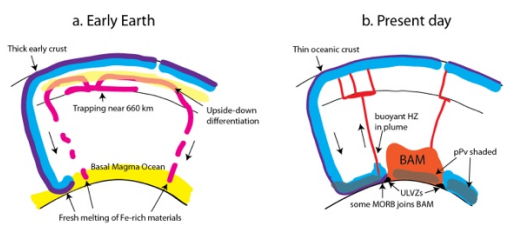

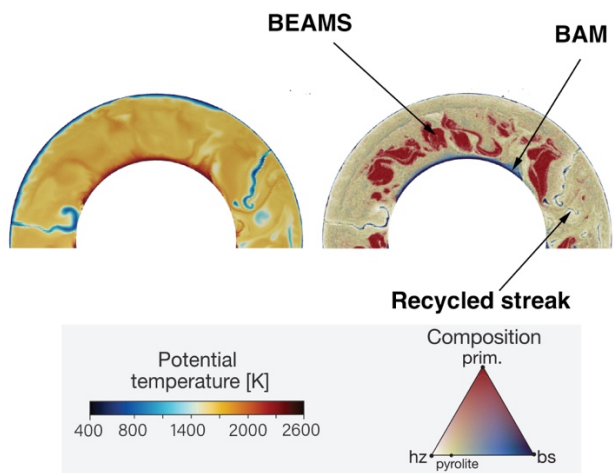

**(c) BEAMS (Ballmer et al., 2017)**

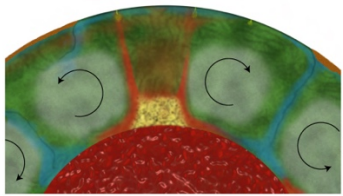

Figure 2. Conceptual models in deep mantle heterogeneity evolved from Archean to present (a) (a) Recycled streak model (Xu et al., 2008), (b) BAM (Tackley, 2012) and (c) BEAMS (Ballmer et al., 2017). (d) Numerical modelling result which includes those conceptual models (Gülcher et al., 2021). Fig. 2(a) is reproduced from Xu et al. (2008) and we are going to obtain permission for use in copyright.

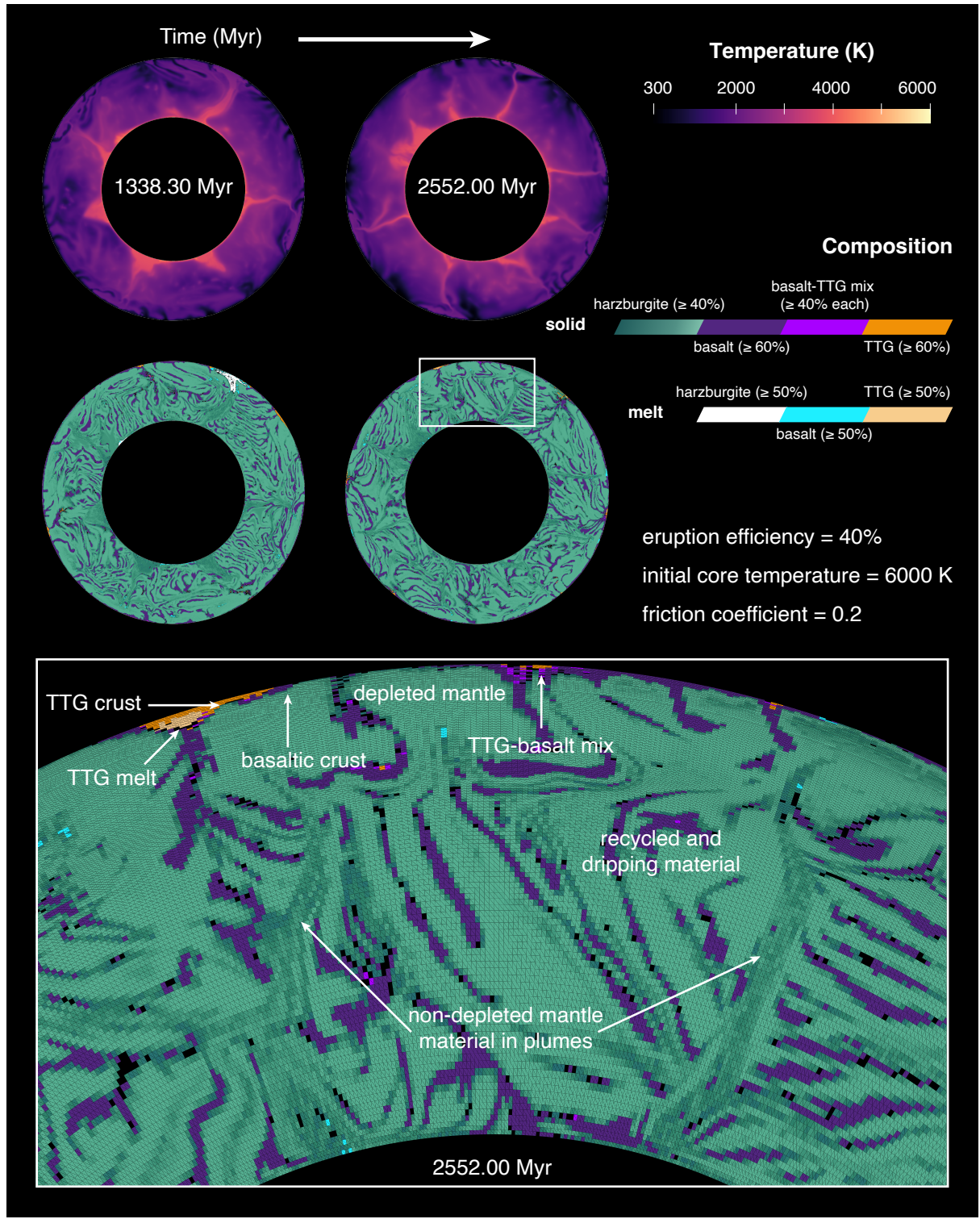

Figure 3. Global-scale modelling on the formation of TTG crust in numerical mantle convection simulations (Jain et al, 2019).

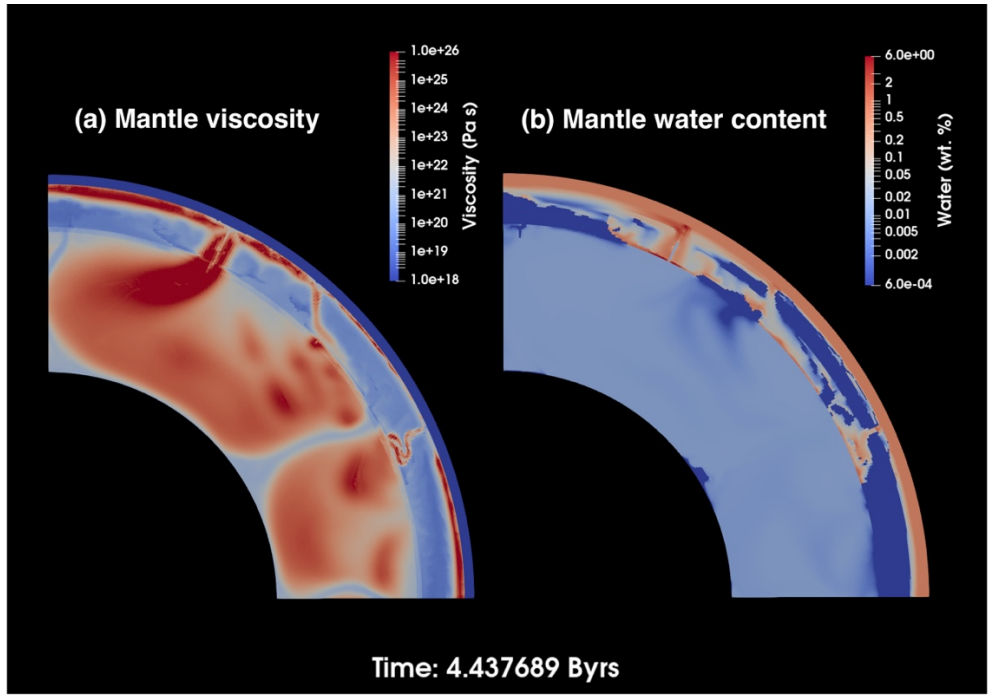

Figure 4. Snapshot of numerical simulation of deep mantle water cycle (a: Viscosity, b: Mantle water content) and (c) mantle water mass and (d) relative sea-level change as a function of time. The detailed procedure in computing the relative sea-level change can be found in Nakagawa (2023).